\documentclass{article}
\usepackage{array,multirow,graphicx}
\usepackage{amsmath}
\usepackage{subcaption}
\usepackage{amsthm}
\newtheorem{lemma}{Lemma}
\providecommand{\keywords}[1]
{
  \small	
  \textbf{\textit{Keywords}} #1
}

\title{Price predictability at ultra-high frequency: Entropy-based randomness test}
\author{Andrey Shternshis$^{1,2}$, Stefano Marmi$^2$\\
\\
        \small 1. Uppsala University, Lägerhyddsvägen 1, Uppsala, Sweden, 75105
        \\
        \small 2. Scuola Normale Superiore, Piazza dei Cavalieri 7, Pisa, Italy, 56126 }
\date{}

\begin{document}
\maketitle
\begin{abstract}
We use the statistical properties of Shannon entropy estimator and Kullback-Leibler divergence to study the predictability of ultra-high frequency financial data. We develop a statistical test for the predictability of a sequence based on empirical frequencies. We show that the degree of randomness grows with the increase of aggregation level in transaction time. We also find that predictable days are usually characterized by high trading activity, i.e., days with unusually high trading volumes and the number of price changes. We find a group of stocks for which predictability is caused by a frequent change of price direction.  We study stylized facts that cause price predictability such as persistence of order signs, autocorrelation of returns, and volatility clustering. We perform multiple testing for sub-intervals of days to identify whether there is predictability at a specific time period during the day.
\end{abstract}


\section{Introduction}

One of the fundamental questions in finance is the predictability of asset prices. Currently, there is a set of tools designed to predict prices in the markets. For instance, traders rely on technical analysis of stocks to make a profit. The prediction power of technical analysis has been the object of many investigations \cite{Lo00, Park07}. Various trading methods can be developed in order to increase profits starting from prediction of future prices which are even only slightly more accurate than a martingale \cite{Hsu16, Hudson21, Akyildirim22, Brasileiro17}.

Randomness of financial data aggregated to daily frequency was investigated in the literature \cite{Molgedey,Dionisio, Ahn}. Studies analyzing the predictability of intraday prices were conducted using hourly \cite{Chu19,Zhang2020} and minute \cite{Rosch16,Calcagnile} frequencies. Prices at millisecond and second frequencies were analyzed in \cite{Leone19}. This research progresses further toward a microscopic examination of financial time series. Ultra-high frequency data is defined as the full record of transactions and their associated characteristics \cite{Engle00}.

This research is dedicated to the predictability of ultra-high frequency data. The high speed of occurrence of new orders makes it difficult to predict the next price before it appears in such a short period of time. Without taking into account transaction costs, we can not ensure that predictability at ultra-high frequency indicates the presence of profitable trading strategies. Moreover, a long memory of price returns at ultra-high frequency is a stylized fact that incorporates predictability into the time series but does not contradict the efficient market hypothesis \cite{Lillo04, Bouchaud03}. We study the relation between other stylized facts and predictability of financial time series at ultra-high frequency. This study complements the work \cite{Calcagnile} on stylized facts that add predictability to time series at a one-minute frequency. We investigate the level of predictability as a function of the length of steps in transaction time. By aggregating by the number of transactions, we increase the time between considered changes in price, dividing days into predictable and unpredictable days. We examine the periods in which the predictability of prices is presented. For instance, we observe for most stocks a high probability of consecutive price directions across several transactions, aligning with the long-memory characteristics of price return signs.

We devise a test for randomness of data starting from the estimation of Shannon entropy. Entropy is defined as an averaged measure of uncertainty about a symbol appearing in a sequence generated by a random source \cite{Shannon}. Maximum uncertainty arises when all symbols from a finite alphabet are independently generated with equal probabilities. A common method for entropy estimation is calculating empirical frequencies of blocks of symbols \cite{Marton}. Previous instances where Shannon entropy served as a measure for assessing randomness in financial time series include \cite{Molgedey, Ahn, Calcagnile}. The entropy of price returns of assets traded on the Moscow Stock Exchange was studied in \cite{Shternshis2}. The time-varying entropy of meme stocks, which experienced sudden price surges driven by social media communities, was investigated in \cite{Shternshis3}. Such instabilities in a market and deviations from a fundamental price value was investigated in \cite{Gardini22}. 

In case of independent and identically distributed symbols, the difference between entropy estimation and its maximum follows $\chi^2$-distribution \cite{Zubkov74}. We make a step forward and use the estimation of relative entropy that allows us to test uncertainty for a symbolic sequence in case when symbols can be not equiprobable. The test statistics is based on all overlapping blocks of symbols and takes into account their dependence. Calculating the statistic is straightforward, requiring only the estimation of frequencies for preselected block lengths. In a financial setting, a significantly high predictability implies that a future price change depends on the price history and thus is not a completely random. Empirical frequencies used in test statistics refer to patterns in sequences of price returns, some of which may be more likely to occur \cite{Shternshis2}.

In a recent work \cite{Brouty23}, randomness of financial prices at daily frequency was tested using conditional probabilities. The authors tested if probabilities of price increase and decrease are the same conditionally on price history. We provide the test for predictability that considers that probabilities that a price moves up or down can be different. 
We test if past price changes are helpful in predicting future price movement. We utilize discretization that distinguishes between positive and negative returns. A test for independence related to permutation entropy of price increments was introduced in \cite{MATILLAGARCIA07}. Considering blocks of $k$ prices, frequencies of $k!$ events need to be calculated. In contrast, our statistic requires the calculation of only $2^k$ frequencies. Approximate entropy was used in \cite{Pincus04, Oh} to measure irregularity of financial time series. The authors fixed the length of blocks $k=3$. We use a larger length for the analysis based on a given sequence length. Entropy of a singular value decomposition was applied to test market efficiency in \cite{Alvarez21}. Monte Carlo simulations were employed by the authors to establish confidence intervals for low entropy values. In our test, we evaluate whether new outputs remain independent of the sequence history, leveraging statistics with known asymptotic distributions. Applying the test based on empirical frequencies of price directions, we get rid of Monte-Carlo simulations.


This article introduces four key contributions. First, we propose the test for randomness of a symbolic sequence. It can be applied with varying numbers of distinct symbols, presuming the sequence length is sufficiently large. Additionally, the individual probabilities associated with each symbol's appearance need not be equal. Second, we investigate the predictability of tick-by-tick data. We show that the degree of predictability decreases when we aggregate simulated and real data by a number of transactions. Third, we investigate the empirical properties of price returns and the difference in the properties of returns for not predictable and predictable time series. Some of the properties are stylized facts of price returns such as fat tails, autocorrelation of returns, and price jumps. Other properties indicative of predictable days encompass heightened daily trading volumes and a substantial number of price changes. We reveal that the probability of consecutive returns displaying the same sign (persistence) serves as a feature for predictable sequences. For a group of stocks, this probability is significantly higher during predictable days. On the other hand, we also observe that the probability of repeating symbols is statistically low for predictable days for another group of stocks. Finally, we localize intervals that we call predictable and find the duration of periods of predictability. We demonstrate that typically there exists a single predictable interval within a predictable day. It's observed that two predictable intervals generally do not occur consecutively. In other words, following an interval with detected predictability, the subsequent interval doesn't display a significant level of predictability. However, considering all transactions of the SNAP stock (Snap Inc.), we are able to detect several predictable time intervals going in a row. Analysis of the characteristics of predictability intervals and their duration for each stock provides insights into the price generation process. 

In Section \ref{Statistical test for the value of entropy}, we propose a statistical test for investigating the predictability of symbols of a sequence. In Section \ref{Tick-by-tick dataset}, we present a chosen group of diverse assets with a set of various characteristics. Section \ref{Predictability of limit order book} is dedicated to applying the predictability test to simulated and real data. Section \ref{Conclusions} concludes the paper.
 

\section{Statistical tests for randomness of symbolic sequence}
\label{Statistical test for the value of entropy}
We introduce a statistical test designed to evaluate the predictability of a sequence. The input for a statistical test is a realization $X$ of a stationary random process $\mathcal{X}$ with symbols from a finite alphabet $A=\{0,1,\cdots,s-1\}$. Symbols of an alphabet with size $s$ can be denoted by integers from 0 to $s-1$ without loss of generality.
$$X=\{x_1,x_2,\dots,x_n\}$$
The goal of the test presented in this section is to verify if symbols of the sequence are independently distributed. Our null and alternative hypotheses are the following.

$\mathcal{H}_0$: The occurrence of a new symbol in sequence $X$ is \textit{independent} of the sequence's prior history.

$\mathcal{H}_a$: Appearance of a new symbol \textit{depends} on past observations of the sequence $X$.

If the null hypothesis is rejected, the probability of guessing the next symbol of a sequence based on the last symbols, is higher than for a random guess. The test is based on empirical frequencies of blocks of symbols introduced by Marton and Shields \cite{Marton}. These empirical frequencies serve to estimate Shannon entropy, which is utilized as a measure of uncertainty. The asymptotic distribution of the Shannon entropy estimation is known and presented in Lemma \ref{Asymptotic distribution of the Shannon entropy}. The asymptotic distribution of the Neyman-Pearson statistics proposed to test predictability is given in Lemma \ref{Asymptotic distribution of the NP-statistics}.

First, we divide a sequence by \textit{non-overlapping blocks} with length $k\in[ 2,n-1]$,

\begin{equation*}
    \hat{X}={\hat{x}_1,\hat{x}_2,\dots,\hat{x}_{n_b}},
\end{equation*}

where $\hat{x}_t=\{x_{(t-1)k+1},x_{(t-1) k+2},\dots,x_{t k}\}$, $n_b=\lfloor\frac{n}{k}\rfloor$. $\hat{x}_t$ represents a number from $0$ to $s^{k}-1$ in the numerical system with base $s$. We calculate $s^{k}$ empirical frequencies $\hat{f}_j$ of blocks from $\hat{X}$:

\begin{equation*}
    \hat{f}_j=\sum_{t=1}^{n_b}I(\hat{x}_t=a_j), a_j\in A^k.
\end{equation*}

Here and further $I$ is an indicator function and $a_j \neq a_l$ when $j \neq l $. The value of $k$ is user-defined. The upper bound of the length  $k$ of symbolic patterns considered to find predictably is validated theoretically and empirically. We take $k=\left[0.5\log_{s}{n}\right]$ as the length of blocks that is proved to be admissible and suggested to use for estimating empirical frequencies in \cite{Shields}\footnote{See Sections 2.3.d and 3.2 or \cite{Marton}}.

Then, entropy estimation is defined as

\begin{equation}\label{entropy estimation}
    \hat{H}=-\sum_{j=0}^{s^{k}-1}\frac{\hat{f}_j}{n_b}\ln{\frac{\hat{f}_j}{n_b}},
\end{equation}
where $\ln()$ is the natural logarithm with the convention $0\ln{0}=0$.

\begin{lemma}[Asymptotic distribution of the Shannon entropy]\label{Asymptotic distribution of the Shannon entropy}
Let's consider a stationary process with symbols from a finite alphabet $A=\{0,1,\cdots,s-1\}$ and its realizations $\{x_t\}_{t=1}^{n}$.  Then, the entropy bias (scaled distance from the maximum of entropy) of a realization defined by Equations \ref{entropy estimation} and \ref{entropy statistics} follows $\chi^2$-distribution with $s^k-1$ degrees of freedom if the probabilities of all blocks of symbols, $\hat{x}_t$, are equal.

\begin{equation}
\label{entropy statistics}
B=2n_b(k\ln{s}-\hat{H})
\end{equation}

\end{lemma}

See \cite{Zubkov74} for the proof. When all probabilities of blocks of symbols are equal, the process is fully unpredictable and the entropy attains its maximum, $k\ln{s}$. We can test unpredictability of a sequence using the entropy bias and the distribution under the null hypothesis $\chi^2_{s^k-1}$. If the probabilities of appearing symbols (blocks with length 1) are not equal, then entropy estimation $\hat{H}$ has asymptotically normal distribution \cite{Zubkov74}. The mean and standard deviation of the normal distribution depend on the entropy value and probabilities of blocks, see \cite{Schurmann96, basharin59}.

Now, we propose a statistics with \textit{overlapping blocks} of symbols that we apply for testing predictability. For the test, we use the Neyman-Pearson (NP) criterion \cite{Neyman28} that contains the likelihood of the sequence under the hypothesis of unpredictability. 
We use \textit{all} blocks with length $k-1$, $k\in[2,n-1]$,

\begin{equation*}\label{sequence}
    \bar{X}={\bar{x}_1,\bar{x}_2,\dots,\bar{x}_{n-k+2}},
\end{equation*}
where $\bar{x}_t=\{x_t,x_{t+1},\dots,x_{t+k-2}\}$.

We construct statistics in Eq.~\ref{statistics} which has $\chi^2$-distribution under $\mathcal{H}_0$. The similar test statistics for Markov chains was considered by Billingsley \cite{Billingsley61}:

\begin{equation}
    \label{statistics}
    D=2\sum_{ij}f_{ij}\ln{\frac{(n-k+1)f_{ij}}{f_{\cdot j}f_{i \cdot}}},
\end{equation}
where 

\begin{equation}\label{empirical freq}
    f_{ij}=\sum_{t=1}^{n-k+1}I\left(\bar{x}_t=a_i\right)I\left( x_{t+k-1}=a_j\right), a_i\in A^{k-1}, a_j \in A
\end{equation}

is the frequency of the event when a block with the last symbol $j \in A$ follows by the block $a_i\in A^{k-1}$ in the sequence $\bar{X}$. We note that $f_{ij}$ are also empirical frequencies of blocks of $k$ symbols. $f_{\cdot j}=\sum_{i}f_{ij}$ and $f_{i \cdot}=\sum_{j}f_{ij}$, $0\le i \le M-1$, where $M=s^{k-1}$ is the amount of blocks of $k-1$ symbols. The convention is $0\ln{0}=0$ and $0\ln{(0/0)}=0$. The asymptotic distribution of the statistics $D$ is described in the Lemma provided below.

\begin{lemma}[Asymptotic distribution of the NP-statistics]\label{Asymptotic distribution of the NP-statistics}
Let's consider a stationary process with symbols from a finite alphabet $A=\{0,1,\cdots,s-1\}$ and its realization $\{x_t\}_{t=1}^{n}$. If the hypothesis of unpredictability, $\mathcal{H}_0$, holds true, then the Neyman-Pearson (NP) statistics (Eq.~\ref{statistics}, \ref{empirical freq}) converges in distribution to $\chi^2$ with $(s^{k-1}-1)(s-1)$ degrees of freedom, i.e.,

\begin{equation*}
    2\sum_{ij}f_{ij}\ln{\frac{(n-k+1)f_{ij}}{f_{\cdot j}f_{i \cdot}}} \xrightarrow[n\to \infty]{d} \chi^2\left((s^{k-1}-1)(s-1)\right).
\end{equation*}

\end{lemma}

We provide the sketch of proof for the distribution of $D$ from Equation \ref{statistics} and subsequently verify it using Q-Q plots in Appendix Section \ref{Q-Q plots}. We show that the proposed test statistics is valid even if the probabilities of appearing symbols are not equal. It is possible because of the denominator in Eq.~\ref{statistics} that mitigates differences in the probabilities of symbols inside the logarithm.
 
Significance of values of entropy estimation and the Neyman-Pearson statistics are defined by $\chi^2$-distributions with suitable degrees of freedom. We calculate p-values associated with values of $B$ and $D$ and degrees of freedom of $\chi^2$-distribution. If a p-value is less than $0.01$, we conclude that the sequence does not exhibit significant predictability with the significance level $\alpha=0.01$. We reject $\mathcal{H}_0$ when p-value$<0.01$. We call intervals where $\mathcal{H}_0$ is rejected \textit{predictable}. While statistics $B$ is based on Shannon entropy estimation, statistics $D$ is scaled Kullback-Leibler (KL) divergence \cite{Kullback} between empirical probabilities $\frac{f_{ij}}{(n-k+1)}$ of blocks of symbols and the probabilities obtained under $\mathcal{H}_0$:

\begin{equation}\label{KL}
\begin{split}
    &\text{KL}\left(\frac{f_{ij}}{(n-k+1)},\frac{f_{\cdot j}f_{i \cdot}}{(n-k+1)^2}\right)=\sum_{ij}\frac{f_{ij}}{n-k+1}\ln{\frac{(n-k+1)f_{ij}}{f_{\cdot j}f_{i \cdot}}}\\
    &=\frac{1}{(n-k+1)}\sum_{ij}f_{ij}\ln{\frac{(n-k+1)f_{ij}}{f_{\cdot j}f_{i \cdot}}}=\frac{D}{2(n-k+1)}.
\end{split}
\end{equation}

Asymptotic distribution of $D$ can be rewritten in form of Gamma distribution, $D \xrightarrow{d} \text{Gamma}\left((s^{k-1}-1)(s-1)/2,2\right)$. Then, KL from Equation \ref{KL} converges as follows,

$$\text{KL}\xrightarrow{d} \text{Gamma}\left(\frac{(s^{k-1}-1)(s-1)}{2},\frac{1}{n-k+1}\right).$$

\section{Dataset}
\label{Tick-by-tick dataset}

\begin{table}[tbh]
\caption{Assets and characteristics of prices}
\centering
\begin{tabular}{|p{3cm}|p{1cm}|p{1.5cm}|p{1.5cm}|p{1.5cm}|p{2.5cm}|p{2.5cm}|}
\hline
Asset           & Ticker & Mean price & Standard deviation of price & Daily trading volume & Daily number of transactions & Average time between transactions  \\ \hline
Apple Inc.             &AAPL        &153.47            &0.93                             &12,184,032                   &136,136                           &0.165                \\ \hline
Microsoft Corporation         &MSFT        &251.78            &1.37                             &4,529,093                    &84,342                           &0.269                \\ \hline
Tesla Inc.             &TSLA        &388.02            &3.81                             &8,686,354                    &178,704                           &0.127               \\ \hline
Intel Corporation             & INTC   &30.15           &0.20                             &7,055,642                    &38,255                          &0.595               \\ \hline
Eli Lilly and Company  & LLY    &327.33            & 1.73                            &370,050                    &11,404                           &2.086                \\ \hline
Snap Inc.         & SNAP   &10.67            &0.14                             &4,967,779                   & 18,521                           &1.358                \\ \hline
Ford Motor Company              & F      &13.93            & 0.10                            &4,468,175                    &12,954                           &1.815                \\ \hline
Carnival Corporation \& plc     & CCL    &9.24            &0.12                             &5,874,376                    &15,372                           &1.518                \\ \hline
SPDR S\&P 500 ETF & SPY    &390.52            & 1.56                            & 9,136,137                    &95,181                           &0.246                \\ \hline
\end{tabular}
\caption*{Mean price, its standard deviation, trading volume, number of transactions, and average time between transactions are calculated for each day and then are averaged over 80 days. Trading volume is summed up for each day. Average time is given in seconds.}
\label{Table: UHF data}
\end{table}

We explore limit order book data \cite{Gould13} downloaded from LOBSTER (www.lobsterdata.com). We consider the executions of visible and hidden orders of a group of assets. The chosen stocks represent diverse industries, differing in characteristics such as average price, volatility, number of transactions, waiting times between transactions, and trading volumes. Additionally, the analysis includes the ETF SPY, designed to track the S\&P 500 Index. The timeframe under consideration spans from 01.08.2022 to 21.11.2022, encompassing a total of 80 trading days. Table \ref{Table: UHF data} presents the tickers of all selected assets along with their respective properties. For each day, the considered daily time interval is from 9:30 to 16:00, that is 390 minutes in total. Times of transactions are recorded with the precision of one nanosecond.

Considering each occurred transaction, we work with data in transaction time \cite{Oomen06}. Discretization is made by distinguishing between positive and negative returns: $0$ corresponds to price decreasing, $1$ corresponds to price increasing. Thus, alphabet $A$ is $\{0,1\}$ and a symbolic sequence is obtained according to binary discretization from Equation~\ref{binary alphabet}:

\begin{equation}
\label{binary alphabet}
\begin{split}
s^{(2)}_t=
\begin{cases}
0, r_t<0, \\
1,r_t>0,
\end{cases}
\end{split}
\end{equation}

where $r_t=\ln{\left(\frac{P_t}{P_{t-1}}\right)}$ are price returns and $P_t$ is the price at time $t$.  All 0-returns are removed.

\section{Predictability of limit order book}
\label{Predictability of limit order book}
\subsection{Modeling ultra-high frequency data}
\label{Modeling long memory of order signs}
\begin{figure}[ht]%
\centering
    \subfloat[\centering predictability of order signs]{{\includegraphics[width=6cm]{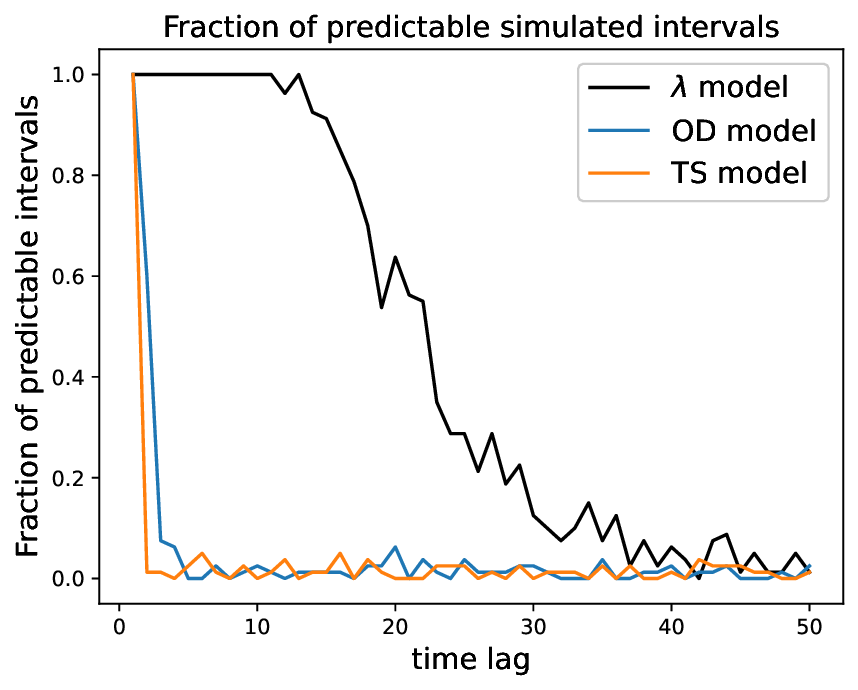} }\label{fig:simulationsigns}}%
    \subfloat[\centering predictability of price returns]{{\includegraphics[width=6cm]{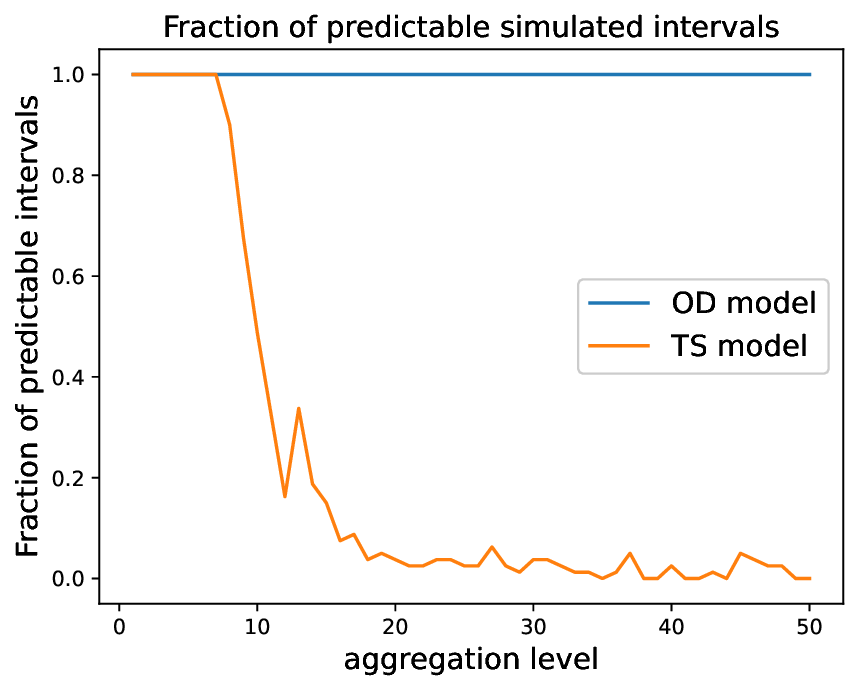} }\label{fig:simulationprices}}%
    \caption{Fraction of predictable days for the $\lambda$, OD, TS model from \cite{Lillo2005, Chiarella02, Bouchaud03}. The x-axis for (a) is the time lag, which signifies the transaction time between order signs. The x-axis for (b) is the aggregation level, denoting the number of transactions considered as one step to compute price returns.}%
    \label{fig:simulations}
\end{figure}

We first apply the proposed method for testing predictability on simulated data. We select models of limit orders exhibiting predictability patterns. We study how the predictability level changes with the aggregation of prices.  Executed orders and trade signs are generated using models that replicate order splitting, herd or chartist behavior, and mean-reverting process \cite{Bouchaud03,Lillo2005,Chiarella02}. According to Kyle's model \cite{Kyle85} including informed traders, a price return linearly depends on the trading volume. In this model, the volume traded by an informed trader depends on the difference between current price and its future fundamental value. According to a behavioral stock market model proposed in \cite{Gardini22}, prices may deviate from their fundamental value due to the behavior of sentiment traders, who buy stocks in rising markets. Positive autocorrelation in order flow was previously observed in \cite{Lillo04} and \cite{Bouchaud03}, where the authors demonstrated that autocorrelation diminishes with increasing aggregation by the number of orders. The $\lambda$ model proposed in \cite{Lillo2005} reproduces the property of the positive autocorrelation decreasing with the larger time lags. The $\lambda$ model introduces a fluctuating number of hidden orders that are divided into equal pieces and submitted gradually. We simulate 80 sequences of signs according to the $\lambda$ model with the length $10^5$ and parameters calibrated in the paper\footnote{The number of hidden orders is $N=21$, the parameter of Pareto distribution describing volumes is $\alpha=1.63$, and the probability of a new hidden order is $\lambda=0.38$.}. We present the fraction of predictable sequences for time lags from 1 to 50 in Figure \ref{fig:simulationsigns} for the $\lambda$ model and two models described below. For small time lags, all sequences are predictable, then the fraction decreases but not monotonically.

Another explanation of the persistence of the signs is herd behavior \cite{Lakonishok92} when traders execute their orders according to the price trend, sometimes against their private information. The model of herd behavior is presented in \cite{Cipriani14}. The parameters of the calibrated model of this behavior suggest that the information obtained by the traders are noisy, which creates uncertainty about the events that occurred. The order driven (OD) model where traders rely on both a fundamental price value and the history of trades is proposed in \cite{Chiarella02}. The predictability for order signs (buy/sell orders) of the OD model\footnote{The parameters are taken from the article \cite{LeBaron07}. These parameters decrease the impact of noisy traders on the deviations of a price from its fundamental value. The authors of \cite{LeBaron07} introduced a modified order flow model incorporating traders' learning and adaptation. It was demonstrated by these authors that price changes within their model display the characteristic of long memory.} drops significantly for time lags larger than 1. However, the predictability of price directions is quite persistent for increasing aggregation level in the amount of transactions as shown in Figure \ref{fig:simulationprices}. An \textit{aggregation level} is a number of transactions taken as one time step. All 80 sequences generated by the OD model are predictable for aggregation levels from 1 to 50 due to the high probability that the price changes the direction in the next time step.

Lastly, we consider the trade superposition (TS) model proposed in \cite{Bouchaud03}. This model posits that each price change results from previous trades, with a specific trade's impact defined by a propagator that diminishes over time\footnote{The model's parameters, sourced from the paper, specify the propagator as $\frac{2.8\times 10^{-3}}{(l+20)^{0.42}}$. The logarithm of volumes follows a normal distribution $N(5.5,1.8)$ and the noise terms have a standard deviation of $0.01$.}. Figures \ref{fig:simulationsigns} and \ref{fig:simulationprices} display the fractions of predictable sequences for order signs and price returns, respectively, based on the TS model. As we will see later in Figure \ref{fig:predictabledays} for real data analysis, the predictability of aggregated prices of frequently traded stocks is similar to the predictability of the TS model.

\subsection{Analysis of real data}
\label{Predictability of Apple's limit order book}
\begin{figure}[ht]%
\centering
    \subfloat[\centering August 1]{{\includegraphics[width=6cm]{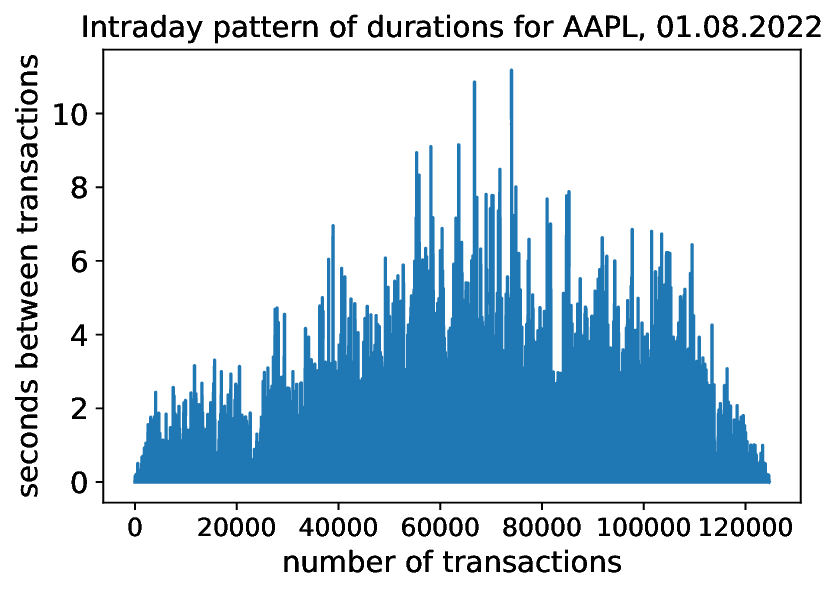} }\label{fig:AAPL0108time}}%
    \subfloat[\centering August 2]{{\includegraphics[width=6cm]{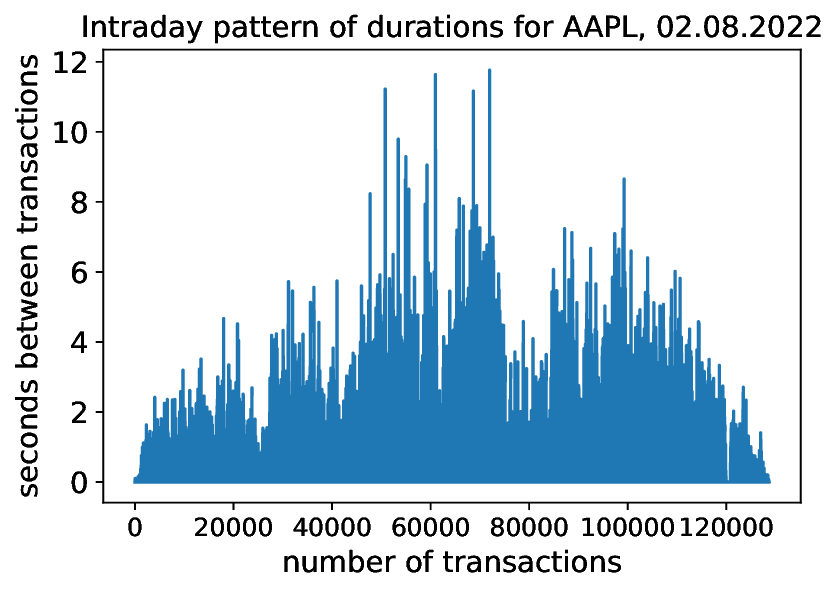} }\label{fig:AAPL0208time}}%
    \caption{Time in seconds between each transaction during two trading days, 01-02.08.2022}%
    \label{fig:AAPLtime}
\end{figure}

Our analysis of real data commences by examining the limit order book of the AAPL stock. We focus on August 1 and August 2 of the year 2022 to demonstrate the behavior of the NP-statistics $D$ and entropy bias $B$ concerning various block lengths. Figure \ref{fig:AAPLtime} illustrates the duration in seconds before each transaction for these two specific days.

\begin{figure}[hbt]%
    \centering
    \subfloat[\centering August 1]{{\includegraphics[width=6cm]{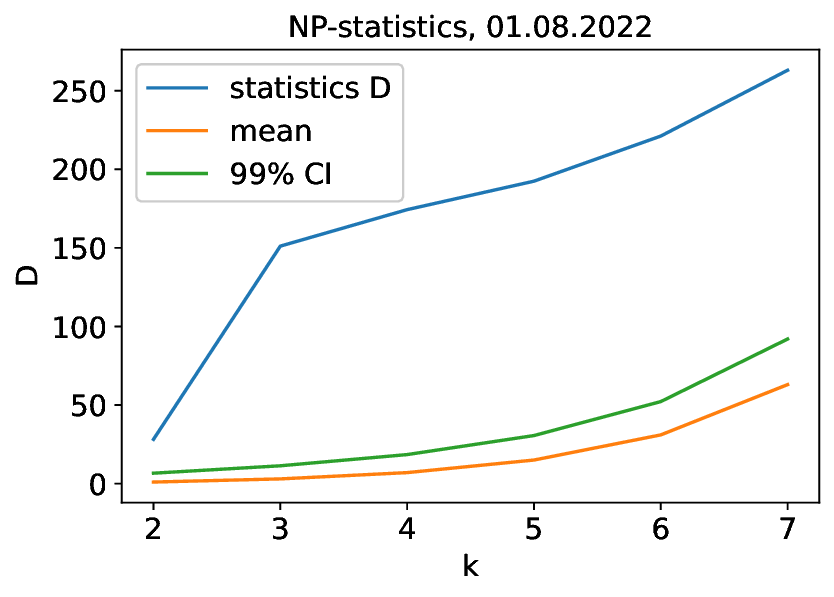} }\label{fig:AAPL0108D}}%
    \subfloat[\centering August 2]{{\includegraphics[width=6cm]{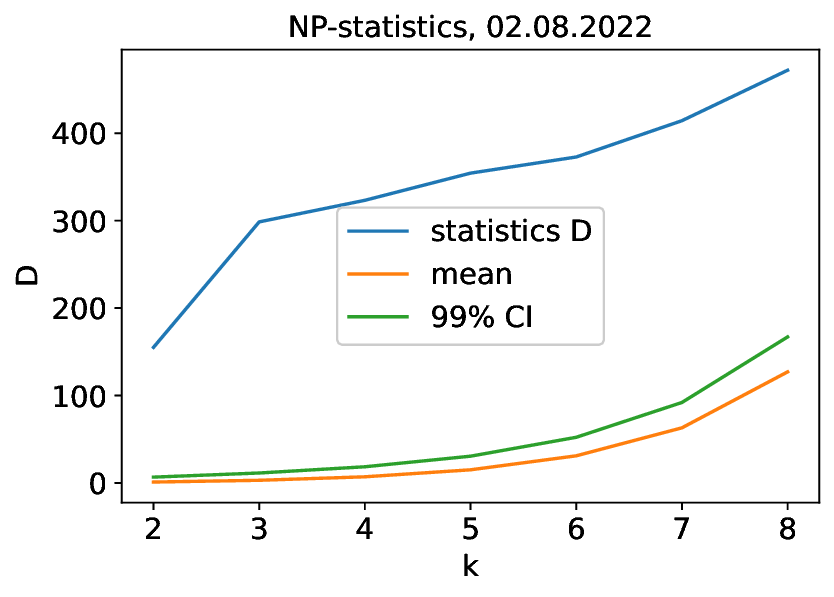} }\label{fig:AAPL0208D}}%
    \caption{NP-statistics for different $k$ for AAPL stock}%
    \label{fig:AAPLD}
\end{figure}

\begin{figure}[hbt]%
\centering
    \subfloat[\centering August 1]{{\includegraphics[width=6cm]{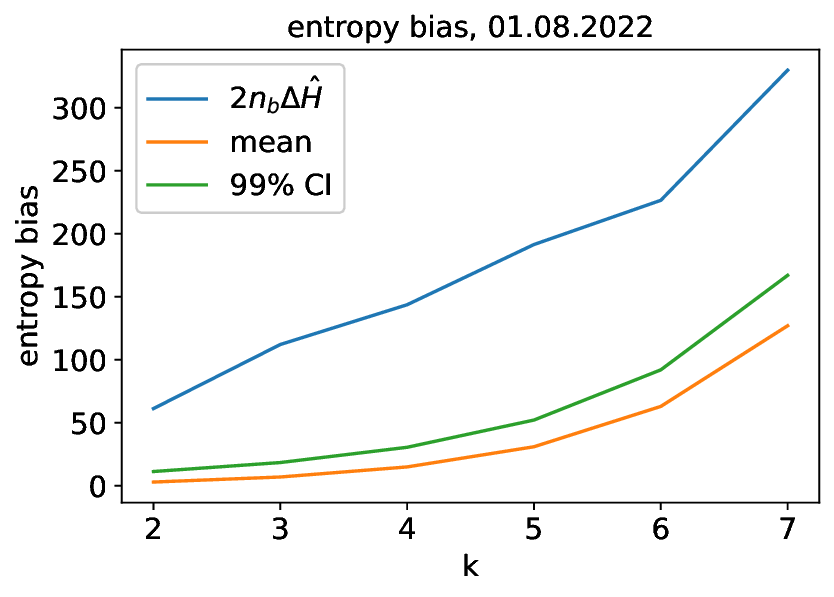} }\label{fig:AAPL0108entropy}}%
    \subfloat[\centering August 2]{{\includegraphics[width=6cm]{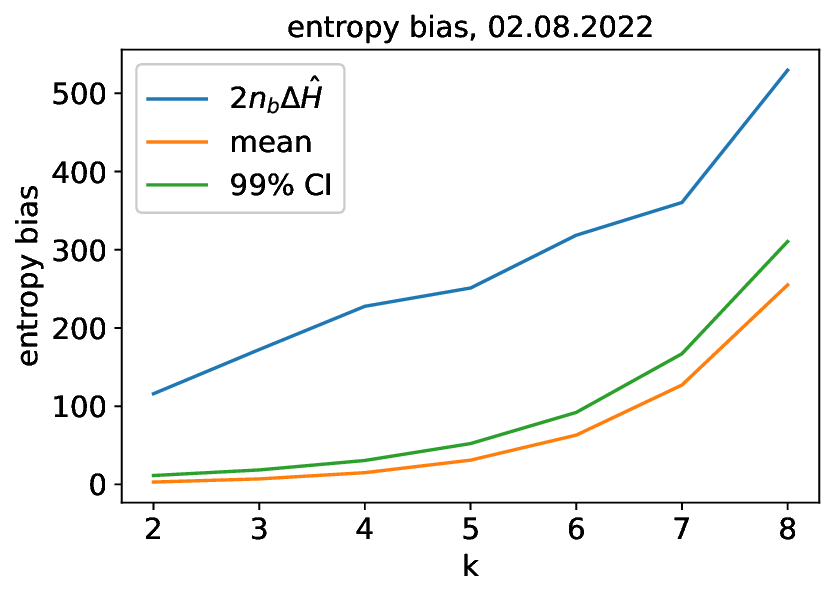} }\label{fig:AAPL0208entropy}}%
    \caption{Entropy bias for different $k$ for AAPL stock}%
    \label{fig:AAPLentropy}
\end{figure}

Figure \ref{fig:AAPLD} illustrates the NP-statistics with 99\% confidence bounds associated with $\mathcal{H}_0$. The mean value of the $\chi^2$-distribution of the NP-statistics under $\mathcal{H}_0$ represents the degrees of freedom. In Figure \ref{fig:AAPLentropy}, entropy bias with corresponding confidence intervals is given. For all values of $k$, sequences are predictable according to the statistics $B$ and $D$. On August 1, the most frequent block with length $2\le k\le 5$ is a sequence of 0s. However, for $k=6,7$ the most frequent block is a repetition of 1. On August 2, the most frequent blocks for corresponding values of $k$ are 00 and 000. When $4\le k\le 8$ the most frequent event is the repetition of $1$ $k$ times. This observation aligns with the well-documented long memory associated with return signs.

\begin{figure}[htb]%
    \centering
    \subfloat[\centering NP-statistics]{{\includegraphics[width=10cm]{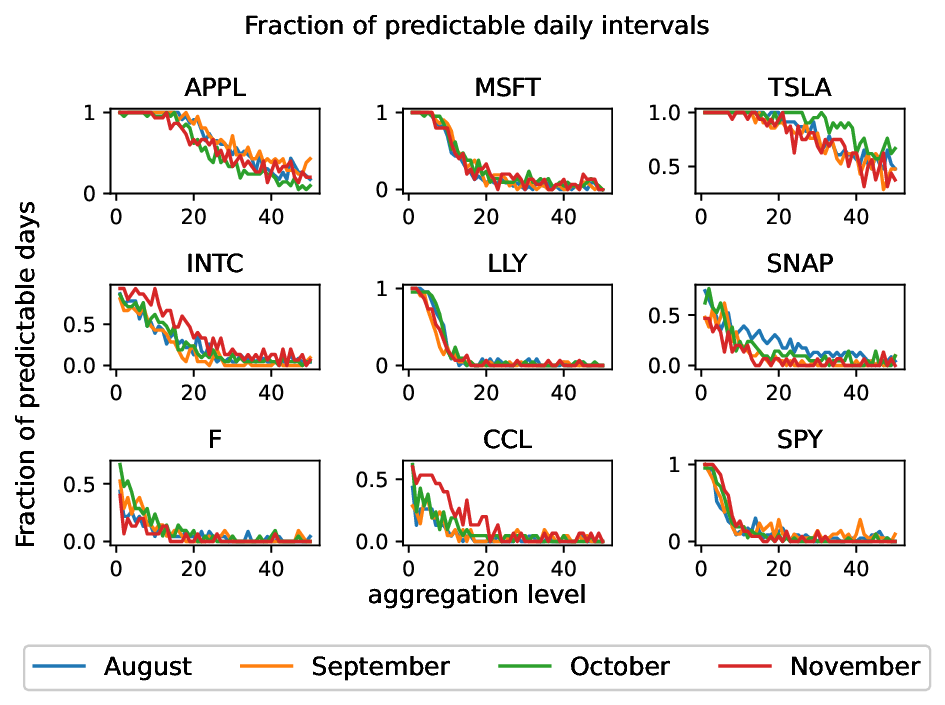}}\label{fig:predictableday_AAPL_chi2}}%
    \qquad
    \subfloat[\centering entropy bias]{{\includegraphics[width=10cm]{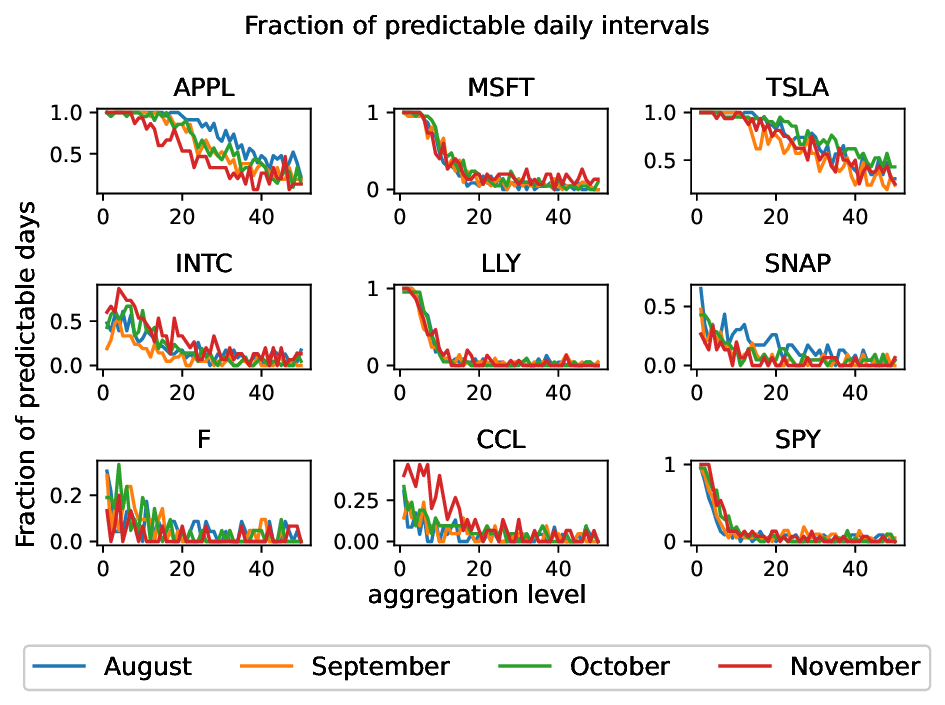}}\label{fig:predictableday_AAPL_entropy}}%
    \caption{Fraction of predictable days for different months and aggregation levels according to (a) NP-statistics and (b) entropy bias}%
    \label{fig:predictabledays}
\end{figure}

Our interest lies in identifying additional factors contributing to predictability beyond symbol repetition due to price persistence. Consequently, we proceed by analyzing aggregated high-frequency data. We use transaction time, that is, we aggregate by a number of transactions. We record the last available price for each time step. We examine the predictability of assets over several months and with different aggregation levels. We plot the fraction of days which is classified as predictable for the four months under consideration in Fig.~\ref{fig:predictabledays}.

There is a noticeable amount of days without predictable patterns even without data aggregation for a group of stocks (INTC, SNAP, F, CCL). As the aggregation level increases, there's a decrease in the fraction of predictable days, although not in a strictly monotonous manner. We couldn't establish a clear correlation between the predictability of assets and the trading months. For instance, a higher level of aggregation is required in August to diminish the predictability of SNAP price. That is, August displays a greater number of days with significant predictability in price returns compared to the autumn months. October stands out as the month with the highest number of predictable days for TSLA. The results obtained from the NP-statistics and entropy bias suggest that the largest fraction of predictable days for INTC and CCL is observed in November.

With a given precision of nanoseconds, some transactions happen at the same time. These simultaneous events might represent scenarios where the volume of a market order surpasses the volume of the best buy or sell limit order. Another scenario involves automatic execution of market orders from different traders at a specific price.  To mitigate the impact of such events on the analysis, we aggregate volumes and consider the final available price for each nanosecond showcasing trading activity.

\begin{figure}[htb]
    \centering
    \includegraphics[width=10cm]{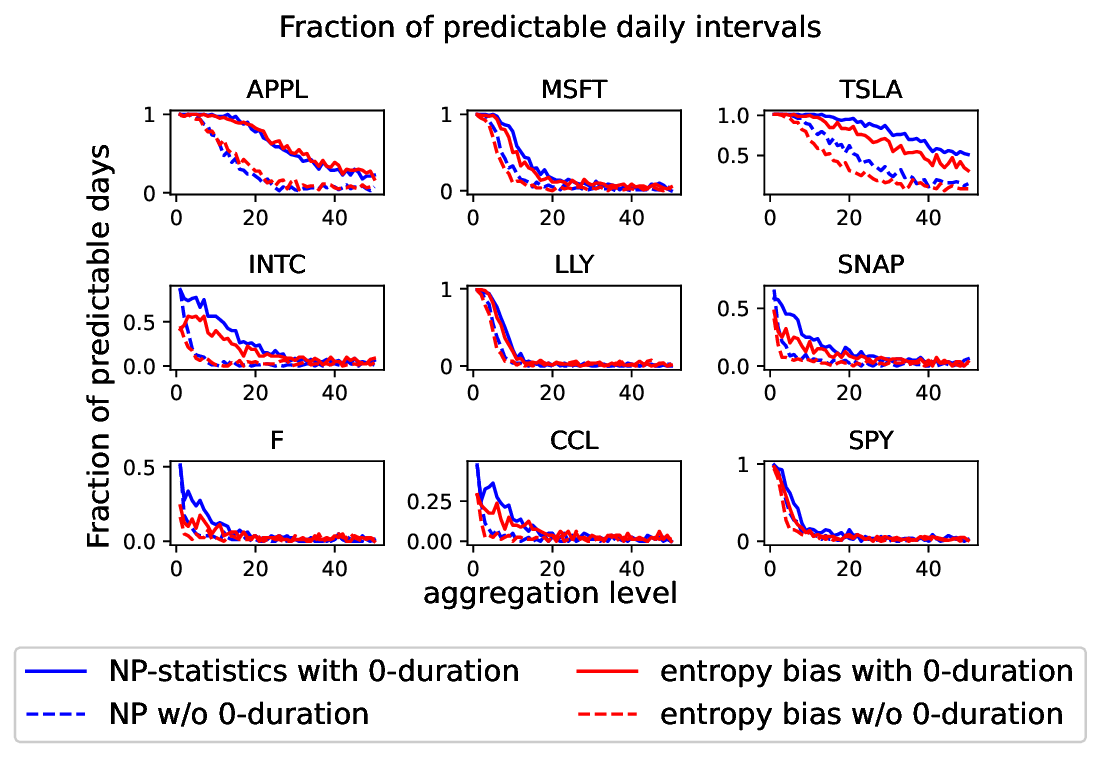}
    \caption{Fraction of predictable days for different assets with and without simultaneous transactions}
    \label{fig:collapse_transactions}
\end{figure}

Figure \ref{fig:collapse_transactions} presents the fraction of predictable days for two statistics and two datasets: full record of transactions and with aggregated transactions in each nanosecond. The exclusion of simultaneous transactions leads to a reduction in the predictability of price return signs.

The empirical findings from Figures \ref{fig:predictabledays} and \ref{fig:collapse_transactions} indicate a noticeable decrease in the degree of predictability as the aggregation level of transactions increases. The decay of predictability over aggregation level fits with results obtained for the simulated data by the TS model by Bouchaud et al.~\cite{Bouchaud03}.

\clearpage
\subsection{Statistics of predictable time intervals}
\label{Statistics of inefficient time intervals}
We show that some assets exhibit unpredictable prices for a part of days. In this section, we explore what price characteristics distinguish days with predictability from others. We first review stylized facts about ultra-high frequency data. 

Lillo and Farmer \cite{Lillo04} conducted a statistical test, concluding that both the signs of market orders and executed limit orders exhibit long-memory processes. They attributed this long memory of order signs to news arrivals and order splitting, which is offset by fluctuations in market liquidity.  Therefore, the long memory does not contradict efficiency of markets when prices incorporate all available information about future values \cite{Fama}. Bouchaud et al.~\cite{Bouchaud02} investigated statistical properties of the limit order books of several stocks. In particular, the authors stated that the distribution of changes in limit order prices exhibits a power-law tail. Engle and Russell \cite{Engle98} noted that the longest duration between transactions appeared in the middle of trading days. According to the authors, clustering of transactions appears because of the size of bid-ask spread and gathering of informed traders. U-shapes of frequencies of large trades, small trades, and market orders were also discovered in \cite{BIAIS95}. In another paper by Engle \cite{Engle00}, he showed that intraday volatility has the similar pattern and attains its minimum in the middle of a trading day. Moreover, significant coefficients of ARMA(1,1) model were detected. Highly dependent microstructure noise was stated by \cite{Christian10}. According to \cite{Epps76}, changes in stock prices between transactions are associated with trading volumes. Some stylized facts including fat tails of price returns, volatility clustering, and leverage effect were discussed in \cite{Bouchaud05}. For a more comprehensive understanding and detailed discussions regarding market and limit orders, we refer to works \cite{Gould13, BIAIS95, Thierry13}.

Now, we consider different parameters of price returns time series and check if there is a dependence between them and predictability. The list of parameters follows.
 
\begin{itemize}
    \item First, we calculate the amount of non-zero returns that is the length of the symbolized sequence $n$ and the fraction of 0-returns. Then, we record lengths of blocks, $k=\left[0.5\log_{s}{n}\right]$. The amount of price returns used as past observations to test the independence is $k-1$.
    \item We compute empirical probabilities of observing blocks with the same symbols $(\hat{p}(0\dots0)+\hat{p}(1\dots1))$ to determine if predictability appears because the price moves in the same direction. We multiply the sum of two empirical probabilities by $2^{k}$ to vanish the difference for different values of $k$.
    \item We calculate $|\hat{p}(1)-\hat{p}(0)|$ to check if predictability is caused by the difference in the amount of price increases and decreases during a trading day. We also write down the magnitude of daily changes in a price to determine if predictability appears when the price significantly changes. For the same purpose, we record the mean value of price returns.
    \item We are interested in autocorrelation with lag 1 of non-zero returns as well as in autocorrelation of their magnitude values. The autocorrelation of magnitudes is a proxy for volatility clustering.
    \item Then, we consider the distribution of price returns. We fit a Student’s distribution of returns distribution and record the degrees of freedom $\nu$, scale, and shift parameters\footnote{Here, we use scipy.stats.t.fit in Python.}. The smaller value of $\nu$, the fatter tails of price returns.
    \item We are interested in whether there is a significant difference in trading volumes.
    \item We also compare the fraction of jumps detected among all price returns. For the detection of jumps, we use the method described by \cite{Lee08}. To employ this test at ultra-high frequency data, we average price returns as suggested by \cite{Christensen14}.\footnote{We use the square root of the amount of price returns as the window size used in the method \cite{Lee08}. The method \cite{Christensen14} requires pre-averaging of price returns. We use the same number of transactions for aggregation and pre-averaging. Jumps are defined with the significance level of $1\%$.}
\end{itemize}

We conduct a statistical test to examine the differences in mean values between predictable and unpredictable days. A p-value below 0.05 indicates a significant difference with a 95\% confidence level. We aggregate prices to collect both sets of predictable and not predictable days. The difference in characteristics may vary depending on the level of aggregation. For example, the persistence of the return sign is more typical for higher frequencies.

In the case of AAPL stock, all 80 days are determined as predictable using the NP-statistics, while 79 predictable days are detected via entropy estimation. To collect days where the hypothesis of unpredictability can not be rejected, we aggregate price returns. We start from the aggregation level $a=15$. Setting the aggregation level to 15 means that we consider only every 15th transaction and use them to build a symbolic series of price increases or decreases. The level of aggregation is chosen to balance between predictable and unpredictable days. Mean values of the specified parameters for AAPL stock are presented in Table \ref{Table: parameters AAPL}, while Table \ref{Table: all parameters} displays the parameter comparison between predictable and unpredictable days for all assets.

\begin{table}[htb]
\centering
\caption{Statistics for predictable and not predictable days of AAPL with $a=15$.}
\begin{tabular}{|p{4.5cm}|p{2cm}|p{2cm}|p{2cm}|}
\hline
 parameter&mean for predictable days  &mean for not predictable days  &p-value  \\ \hline
Sample size (number of days) &35  &  45 &  \\ \hline
number of non-zero returns &5225 &  4370  &0.009**  \\ \hline
fraction of 0-returns &0.142 &  0.138 &0.320  \\ \hline
$k$ &6 &  5.889  &0.118 \\ \hline
$(\hat{p}(0\dots0)+\hat{p}(1\dots1))2^k$ &3.084  &2.635  &3.152$\times10^{-5}$**  \\ \hline
$|\hat{p}(1)-\hat{p}(0)|$ &0.051  &0.050  &0.874  \\ \hline
magnitude of daily log-price increment  &0.017  & 0.012 &0.093 \\ \hline
mean price returns  &2.185$\times10^{-7}$ & -6.673$\times10^{-7}$ &0.315  \\ \hline
magnitude of autocorrelation of non-zero returns & 0.044 & 0.027 &0.002**  \\ \hline
magnitude of autocorrelation of absolute values& 0.115  & 0.088 &0.007** \\ \hline
$\nu$ of t-distribution &2.220  & 2.694 &0.31 \\ \hline
scale of t-distribution &1.367$\times10^{-4}$  & 1.464$\times10^{-4}$  &0.195 \\ \hline
magnitude of shift of t-distribution &9.082$\times10^{-6}$  & 9.991$\times10^{-6}$  &0.519  \\ \hline
daily volume  &13,367,901  & 11,257,199&0.026*  \\ \hline
fraction of jumps  &1.934$\times10^{-4}$ & 1.367$\times10^{-4}$ &0.262  \\ \hline
\end{tabular}
\caption*{In the last column, * is rejection of equal means with 0.05 significance and ** stands for 0.01 significance.}
\label{Table: parameters AAPL}
\end{table}

\begin{table}[htb]
\centering
\caption{\footnotesize Statistics for predictable and not predictable days}
\footnotesize
\begin{tabular}{|p{4.5cm}|l|l|l|l|l|l|l|l|l|l|}
\hline
 parameter &\multicolumn{1}{c}{AAPL}&  &MSFT  &TSLA  &INTC &LLY &SNAP &F &CCL &SPY\\ \hline
aggregation  level &10 &30  &10  &25  &2 &5 &1 &1 &1 &5\\ \hline
number of predictable days &35 &3 &26  &35  &44 &44 &53 &41 &38 &29\\ \hline
number of non-zero returns &$>$** &   & $>$**  &$>$*  &$>$** &$>$** &$>$** &$>$** &$>$* &$>$*\\ \hline
fraction of 0-returns & &   &  &  & & &$>$* &$>$** &$>$** &\\ \hline
$k$ &  &$>$**  &  &  & &$>$* & &$>$** &$>$* &\\ \hline
$(\hat{p}(0\dots0)+\hat{p}(1\dots1))2^k$ &$>$**  &  & $>$**  &$>$**  &$>$** &$>$** & & & &$>$**\\ \hline
magnitude of daily log-price increment &  &  &  &  & & & &$>$** & &\\ \hline
mean price returns   &  &  &  &  & & &$>$* & &$>$* &\\ \hline
magnitude of autocorrelation of non-zero returns &$>$**  & &  &$>$**  &$>$** &$>$** &$>$** &$>$** &$>$** &$>$**\\ \hline
magnitude of autocorrelation of absolute values & $>$** &$<$**   &  &  & & &$>$** &$>$* &$>$** &\\ \hline
$\nu$ of t-distribution &  &$<$*  &  &  & & & &$>$* &$>$** &$<$*\\ \hline
scale of t-distribution &  &$<$*  &  &  & &$<$* &$<$* & & &\\ \hline
daily volume  & $>$*  &  & $>$**  &  &$>$* &$>$** &$>$** &$>$** &$>$** &$>$*\\ \hline
fraction of jumps  &  &$<$**  &  &  & & & & &$>$* &\\ \hline
\end{tabular}
\caption*{\footnotesize  * is rejection of equal means with 0.05 significance and ** stands for 0.01 significance, $>$ stands for larger mean for predictable days, $<$ stands for larger mean for not predictable days. There is no significant difference found for parameters $|\hat{p}(1)-\hat{p}(0)|$ and magnitude of shift of t-distribution. For AAPL stock, the results for two aggregation levels are shown.}
\label{Table: all parameters}
\end{table}

Predictable days exhibit significantly higher amounts of price movement and trading volumes. Additionally, autocorrelation values for non-zero returns and their magnitudes are notably higher on predictable days. Moreover, the probability of repeating the same symbol is significantly elevated during days with predictable price returns. To mitigate predictability resulting from the persistence of signs, we increase the aggregation level. With an aggregation level of $a=30$ only three days (02.08, 20.09, 05.10) are identified as predictable. 

Then, we divide days into two groups after aggregating by $a=30$ using the test-statistics  $B$ derived from entropy estimation from Eq.~\ref{entropy statistics}. There are 14 predictable days, the results are given in Table \ref{Table: parameters AAPL with B} in Appendix Section \ref{Appendix}. Even after aggregation by $30$ transactions, the predictable days have a higher probability of repeating the direction of the price. They are also characterized by the larger difference in frequencies of increasing and decreasing of price. To mitigate the difference, the median of price returns could be used instead of zero during the discretization process by Eq.~\ref{binary alphabet}. Further, we use NP-statistics $D$ that takes into consideration the difference between frequencies of symbols by design.

For the Microsoft stock, we discover 26 predictable days after aggregating data by $a=10$ transactions. According to the test for the difference in mean value of $(\hat{p}(0\dots0)+\hat{p}(1\dots1))2^k$, the price direction is more persistent for predictable days. With a slightly larger aggregation level of $a=15$, this difference is eliminated and only four predictable days remain. These days are listed in Table \ref{optimal partition all} in Appendix Section \ref{Appendix}.

For the stock TSLA, we identify 35 predictable days with aggregation $a=25$. On these days, we observe a higher number of price changes, a stronger autocorrelation of non-zero returns, and increased probabilities of blocks repeating a symbol. Increasing the aggregation level by 5, we observe that the probability of repeating a symbol is significantly high even when $a=60$. However, by further increasing the aggregation level to $a=65$, only 4 predictable days remained (05.08, 14.09, 15.09, 08.11), and the previously significant differences disappeared. Regarding the influence of news on price predictability, it's worth noting that on August 5, Tesla shareholders approved a 3-1 stock split. On September 14, Tesla faced a lawsuit for false advertising of its autopilot technology. Additionally, on November 8, Tesla recalled over 40 thousand cars\footnote{The news are taken from edition.cnn.com and reuters.com.}. These instances of significant events or news might be associated with predictable price behavior at high frequencies. Traders tend to react more actively to new information, leading to increased trading activity reflected in price changes and higher trading volumes. In general, high trading activity observed during predictable days across all considered assets suggests a possible connection between predictability and the arrival of news or significant events. For a more in-depth exploration of how prices react to news, we refer to review \cite{Corrado11}.

For the stock INTC, we identify 44 predictable days using the aggregation level of $a=2$ transactions. These days notably differ in terms of daily changes and autocorrelation of returns. However, as we increase the aggregation level to $a=7$ transactions, only 8 predictable days remained, characterized by significantly higher amounts of non-zero returns. At an aggregation level of $a=10$, the only predictable day is October 13. For the stock LLY, the set of different characteristics is similar to the case of INTC when $a=5$. At a higher aggregation level of $a=10$ transactions, only three predictable days (03.08, 29.09, 28.10) are found with significantly high probability of repeating symbols.

For the stock of Snapchat, we find 53 predictable days without employing aggregation. These predictable days do not exhibit a large probability of repeating symbols. The distinguishing features of predictable days are a larger fraction of 0-returns, an increased mean return, a high autocorrelation of returns and their absolute values, and a smaller scale of fitted t-distribution. The case of the stock CCL is the only one where the percentage of price jumps is significant and higher for predictable days. We present the results for the stock CCL in Table \ref{Table: parameters CCL} in Appendix Section \ref{Appendix}. Last, we consider the price of the ETF SPY with $a=5$ in Table \ref{Table: parameters SPY} in Appendix Section \ref{Appendix}. We discover that price returns of predictable days have fatter tails than returns of unpredictable days. Using the aggregation level $a=10$, we remain with 9 predictable days presented in Table \ref{optimal partition all} with no significant difference in probabilities of repeating symbols.

\subsection{Predictability of pairs of signs}
\label{Predictability of pairs of orders}
Here, we explore the predictability of pairs of price changes, that is the randomness of future price movement observing the last price change. We set the user-defined parameter $k$ to be equal to 2. For the majority of considered assets, we find that predictability is associated with a high frequency of blocks with the same price direction. However, for three stocks, SNAP, F, CCL, we fail to detect the high frequency of blocks $0\dots0$ and $1\dots1$ without aggregation of prices. Here, we consider the three stocks to test if a price direction depends on the previous recorded decrease or increase. In other words, we test $\mathcal{H}_0$ setting $k=2$ and estimate $p(00)+p(11)$ for predictable and not predictable days. A summary table is Table \ref{changing pattern}. For all three stocks we detect a significant difference in $\hat{p}(00)+\hat{p}(11)$ for the two groups of days. However, the sum of two probabilities is less than $0.5$ for predictable days. That is, the probability of changing price direction is significantly high for predictable days. Therefore, predictable days are more likely to have a pattern of changing symbols indicating an increase or decrease in price.

\begin{table}[htb]
\caption{Probabilities of repeating symbols for predictable and not predictable days}
\begin{tabular}{|p{1cm}|p{2.5cm}|p{2.5cm}|p{3cm}|p{3cm}|}
\hline
Stock & N. of predictable days & $\hat{p}(00)+\hat{p}(11)$ for predictable days & $\hat{p}(00)+\hat{p}(11)$ for unpredictable days & p-value for difference in mean values \\ \hline
SNAP  &  36                     & 0.469                                                 &   0.501                                               &  0.0014                                    \\ \hline
F & 25     & 0.460                       & 0.493                                               &   2.671$\times 10^{-10}$                                                                                    \\ \hline
CCL   & 27                       &      0.478                                          &   0.5                                               &   0.044                                   \\ \hline
\end{tabular}
\label{changing pattern}
\end{table}

The number of predictable days is fewer when considering $k=2$  compared to the value of $k=\left[0.5\log_{s}{n}\right]$, where $n$ represents the length of a symbolic sequence. As the number of symbols used to test dependence on past history increases, more days exhibit predictability. The probability of repeating the same symbol is approximately 0.5 for unpredictable days, whereas it is notably smaller for predictable days. However, this characteristic of predictable days for the three stocks diminishes as $k$ increases to around 5 or 6 as indicated in Table \ref{Table: parameters CCL}. That is, a pattern of switching direction with each price movement tends to last for less than 5 price changes on average.

\subsection{Localization of predictable intervals}
\label{Localization of inefficient intervals}

In this section, we consider the length of the interval used to detect predictability. In previous sections, we investigate daily time intervals. The key question here is whether there exists a smaller time interval within a predictable day where significant predictability can be identified. The motivation for searching for the smaller interval is to localize the period when price predictability occurs. Additionally, using a smaller time interval necessitates less historical data for computing entropy-based estimations. Therefore, employing a smaller rolling window enables quicker detection of price predictability.

Since rolling window inside a trading day implies multiple tests, the Šidák correction \cite{Sidak67} of the significance level is used. Moreover, to ensure independence among the conducted tests, non-overlapping intervals within a trading day are considered. The maximum value of considered partitions is $S_{max}=\lfloor(n-k+1)/1000\rfloor$, so that $1000$ is the minimum length of intervals. For each trading day, we aim to detect predictability for at least 1 from $S$ partitions with significance level $1-0.99^{1/S}$. We record the maximum value of such $S\le S_{max}$. Smaller intervals require a smaller $k$ for analysis.  We present the results in Table \ref{optimal partition all} in Appendix Section \ref{Appendix}.

There are three days for AAPL stock where predictability is detected when the data is aggregated. For all three days predictability is detected only for daily time intervals ($S=1$). For the MSFT stock, there are three days where predictability is detected only for the part of the day ($S>1$). Regularity patterns are detected at the end of the day for August 3 and October 26. For August 5, predictability is presented for the first half of the day. It disappears at the next subsequent non-overlapping interval. For each predictable day of the ETF SPY, there is only one from $S$ intervals where predictability is found. Predictability disappears from the time interval to the next subsequent non-overlapping interval. For instance, predictability was noticeable in the middle of the trading day on August 17. Regarding TSLA stock, detectable predictability was identified in both halves of the day on August 5, indicating the potential to reduce the sequence length for predictability analysis by half and still observe significant patterns throughout the day. For 13 out of 30 days where $S_{max}>1$ for the Ford stock, the predictability can be found in the second half of a day. Similarly, there are 14 days in the sample of 31 days with $S_{max}>1$ where the predictability is detected only in the second part of a trading day for CCL.

For the stock SNAP, we detect days when predictable time intervals occur several times during a day. We notice that some predictable intervals cluster together and highlight such intervals in the Table \ref{optimal partition all}. For instance, we observe 7 predictable intervals going in a row on October 21 and 24. However, we do not find evidence that predictive intervals follow each other for the other stocks considered.

\section{Discussion and Conclusions}
\label{Conclusions}

We have applied a statistical test for the randomness of a symbolized sequence. A short summary of the method for detecting predictability is given below.

\begin{itemize}
     \item First, we estimate the Shannon entropy using empirical frequencies of blocks of symbols as suggested in \cite{Marton}. Using empirical frequencies obtained by rolling a window with a certain length, we calculate the NP-statistics.
    \item The NP-statistics has $\chi^2$-distribution according to \cite{Billingsley61} and \cite{Wilks35}. We have found degrees of freedom of the $\chi^2$-distribution that depends on the length of blocks and the size of the alphabet.
    \item The statistics is a scaled KL-divergence \cite{Kullback} that measures difference between empirical probabilities and theoretical probabilities under the null hypothesis of unpredictability.
    \item The method is computationally fast since it requires only the empirical frequencies of blocks of symbols.
\end{itemize}

We have studied the predictability of asset prices at ultra-high frequency. Considering signs of price changes, we construct binary sequences for all recorded executed transactions. The signs of trades have a long memory in such a microscopic view of transaction data \cite{Lillo04, Bouchaud03}. We have shown that the degree of predictability decreases with the increase of aggregation level.  We apply aggregation by the number of transactions and work with transaction time. Transaction times have an uneven time intervals in seconds, which has been empirically shown by \cite{Engle98, BIAIS95}. We have shown that the significant predictability level decreasing with larger aggregation level can be explained by splitting hidden orders into pieces as modeled by \cite{Lillo2005} or by mean reverting as modeled in \cite{Bouchaud03}. According to \cite{Lillo04, Bouchaud03}, such type of predictability does not lead to arbitrage opportunities because it is compensated by fluctuations in transaction costs and liquidity and by interaction between market makers and informed traders. We have also demonstrated that transactions appearing simultaneously with the precision of a nanosecond contribute to a larger predictability level.

We apply the correction proposed in \cite{Sidak67} to make multiple tests for predictability for short intervals during predictable days. In most cases, a single predictable interval is identifiable by partitioning a day into uniform intervals based on transaction time. In such a way, we determine both the position of this interval relative to the time of day and its duration. For the stock SNAP, we have found several groups of such predictable intervals following each other.

Applying the test, we distinguish predictable days from not predictable days. We have shown that the probability that the price of an asset has several subsequent movements in the same direction is one of the factors affecting the predictability of the prices. For a group of assets, predictable days are characterized by repeating signs of price returns. Except LLY, trades on these assets appear at extremely high frequencies, i.e., less than one second on average. The repetition of price direction is explained by the appearance of news, the reaction to them, and the splitting of one order into parts \cite{Busse2002, CHAN95}. Conversely, another group of stocks (SNAP, F, CCL) demonstrated a lower percentage of predictable days before aggregation. In these cases, days featuring predictable time intervals were characterized by a significantly reduced probability of price moving in the same direction twice. This group differs by relatively low prices and their standard deviations. The pattern of changing price direction can be explained by a bid-ask bounce and fluctuations of the price around a low mean value. We presume that the occurrence of this behavior depends on the frequency of transactions.

For 8 out of 9 assets under consideration, non-zero price returns during predictable days have high autocorrelation. Highly significant coefficients of an AR model for tick-by-tick data were empirically investigated by \cite{Engle00} and \cite{Christian10}. Some stylized facts of price returns at ultra-high frequency data including fat tails of return distribution and volatility clustering were discussed in \cite{Bouchaud05}. To explore fat tails of price returns, we estimate degrees of freedom of the fitted t-distribution of the price returns. For the ETF SPY, we discover that price returns of predictable days have fatter tails than returns recorded during not predictable days. However, we have the opposite result for the stocks of Ford and Carnival Corporation, where not predictable days are described by price returns with fatter tails. We check volatility clustering by measuring the autocorrelation of the absolute values of returns. The autocorrelation is significantly greater during predictable days for the stocks AAPL, SNAP, F, and CCL.

We notice that predictable days of AAPL, MSFT, INTC, LLY, F, CCL, and SPY are characterized by larger trading volumes and the larger amount of non-zero price changes in comparison with not predictable days. The dependence between price changes and trading volumes was described in the paper of Kyle \cite{Kyle85}. The distinction in the volumes and price changes suggests a correlation between predictability and trading activity, particularly in response to market news events. This assumption aligns with existing research demonstrating the influence of news on stock prices. Stock prices react to announcements about stock dividends and splits \cite{Grinblatt84}. Weekly price returns react to attention to news and their tone \cite{Huynh17}. Public news affect monthly price returns \cite{Chan03}.

We have presented the approach for testing predictability of financial data. Using this approach, we find days with a statistically significant level of predictability. Aggregating the data so that the time between the transactions under consideration increases, we allocate a smaller group of days with predictability. Various methods exist for aggregating data, including calendar time, transaction time, and tick time.  Alternative approaches involve counting volumes traded and price changes by a certain amount. A direction for future exploration involves comparing the predictability observed in data aggregated through diverse methodologies.

\appendix
\section{Asymptotic distribution of the Neyman-Pearson statistics}
\label{Q-Q plots}

\begin{figure}[htb]
    \centering
    \includegraphics[scale=0.4]{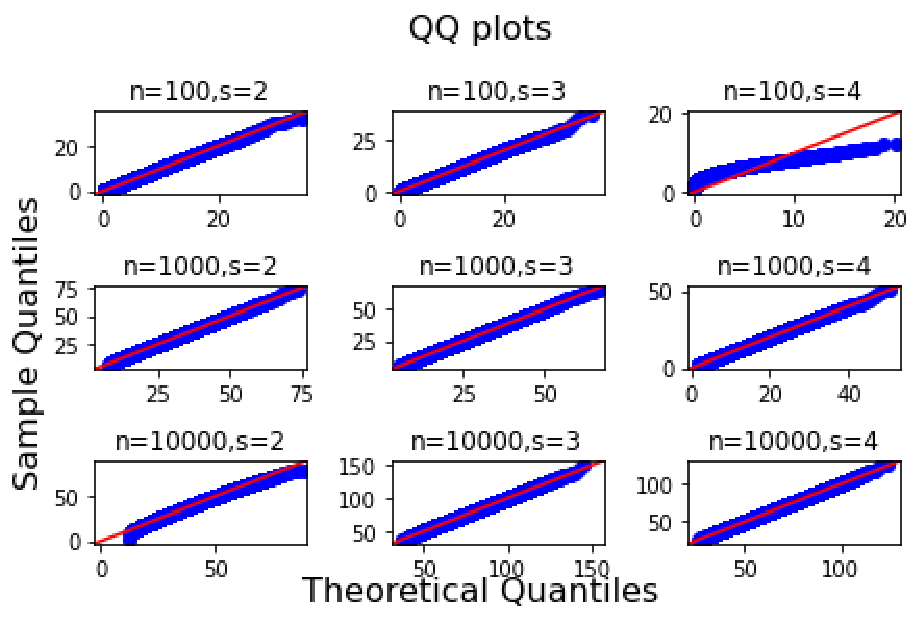}
    \caption{QQ plot for entropy bias when probabilities of symbols are equal}
    \label{fig:QQentropy}
\end{figure}

We outline the proof of Lemma \ref{Asymptotic distribution of the NP-statistics}. The lemma gives the asymptotic distribution of the statistics $D$ in Equation \ref{statistics} based on empirical frequencies of blocks of symbols of a sequence. If symbols of sequence $X=\{x_t\}_{t=1}^{n}$ are independently distributed, then the sequence $\bar{X}$ of blocks  $\bar{x}_t=\{x_t,x_{t+1},\dots,x_{t+k-2}\}$ follows the first order Markov process. The transition matrix of the Markov chain has dimension $M\times M$, where $M=s^{k-1}$. However, there are only $s$ non-zero probabilities whose sum is equal to one in each row of the matrix. Thus, the transition matrix can be converted into $M\times s$ matrix with only positive entries. This transformation is feasible since an output $\bar{x}_t$ in sequence $\bar{X}$ differs from $\bar{x}_{t-1}$  solely by a single new symbol $x_{t+k-2} \in A$.

We denote by $p_{ij}$ transition probability from the block of symbols $i \in \{0,1,\dots,M-1\}$ to the block with the last symbol $j\in A$. The null hypothesis $\mathcal{H}_0$ can be restated as $p_{ij}=p_j$, indicating that the probability of obtaining a new symbol does not rely on the previous $k-1$ symbols. Thus, the hypothesis $\mathcal{H}_0$ assumes the transition matrix with the following structure.

\begin{equation*}
   T= \begin{bmatrix}
    p_{0} & p_{1}  & \dots  & p_{s-1} \\
    p_{0} & p_{1}  & \dots  & p_{s-1} \\
    \vdots & \vdots &\ddots  & \vdots  \\ 
    p_{0} & p_{1} & \dots  & p_{s-1}
\end{bmatrix}
\end{equation*}

The asymptotic distribution of the Neyman-Pearson statistics (Eq.~\ref{statistics}) is $\chi^2$-distribution that can be shown using a characteristic function \cite{Wilks35}. There are $s$ probabilities with the sum of 1 in each of $M$ rows in the matrix $T$ and it contains $s$ parameters with one constraint. Therefore, the degrees of freedom of the NP-statistics are $M(s-1)-(s-1)=(M-1)(s-1)$ according to Theorem 4.1 of \cite{Billingsley61}.

Then, we empirically show that the entropy bias (Eq.~\ref{entropy statistics}) and the NP-statistics (Eq.~\ref{statistics}) follow $\chi^2$-distribution with $s^k-1$ and $(s^{k-1}-1)(s-1)$, respectively. As defined in the main text, $s$ is the size of the alphabet and $k$ is the length of blocks. We consider alphabets with 2,3,4 symbols and take sequences with three different lengths, $\log_{10}n=2,3,4$. The length of blocks depends on the sequence length, $k=\left[0.5\log_{s}{n}\right]$. For each plot we simulate $N=10^{5}$ sequences. We provide the QQ-plot for entropy bias in Figure \ref{fig:QQentropy}.

\begin{figure}[htb]
    \centering
    \includegraphics[scale=0.4]{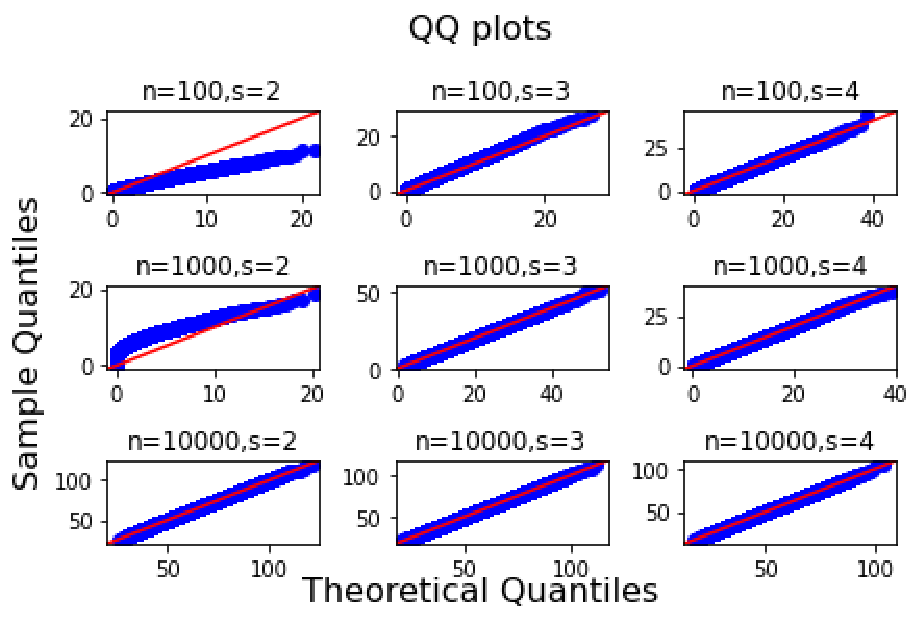}
    \caption{QQ plot for NP-statistics when probabilities of symbols are equal}
    \label{fig:QQ1}
\end{figure}

\begin{figure}[htb]
    \centering
    \includegraphics[scale=0.4]{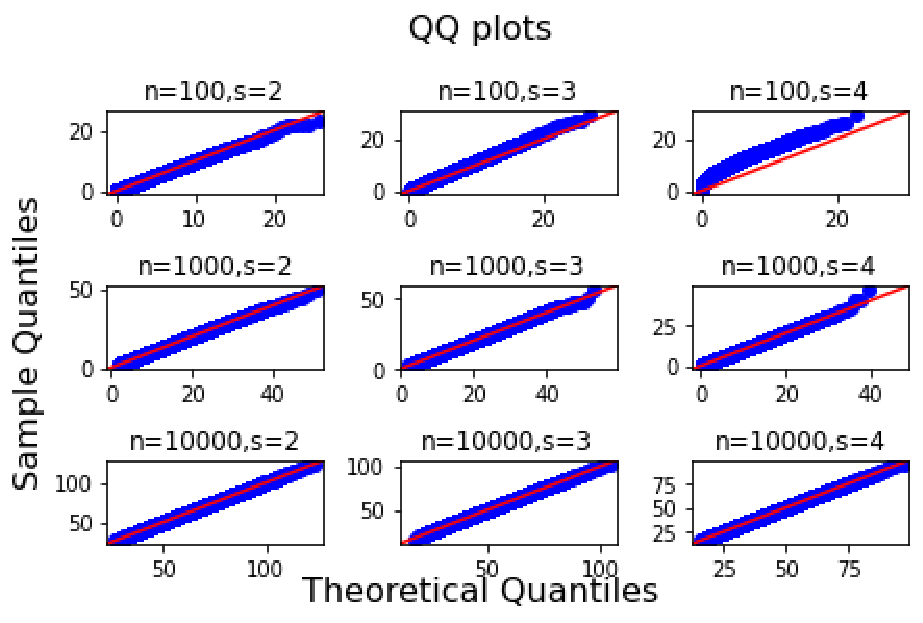}
    \caption{QQ plot for NP-statistics when probability of symbol 0 is twice that of another symbol}
    \label{fig:QQ2}
\end{figure}

The next two Figures show the QQ plots for calculated NP-statistics. For Figure \ref{fig:QQ1}, sequences have equal probabilities of appearing symbols. In contrast, Figure \ref{fig:QQ2} displays the scenario where the probability of symbol 0 is twice as great as the probabilities of other symbols. These figures demonstrate the convergence of the empirical distribution of $D$ to the theoretical $\chi^2$-distribution.

\section{Statistics and partitions of predictable days}
\label{Appendix}

This section comprises results pertaining to statistics derived from predictable and not predictable days.  The results for the stock AAPL using entropy bias are in Table \ref{Table: parameters AAPL with B}. Detailed results for the stock CCL and ETF SPY are provided in Tables \ref{Table: parameters CCL} and \ref{Table: parameters SPY}, respectively. Finally, Table \ref{optimal partition all} shows partitions based on the test of randomness conducted for intervals of predictable days.

\begin{table}[htb]
\centering
\caption{\footnotesize Statistics for predictable and not predictable days of AAPL determined by entropy bias  with $a=30$.}
\footnotesize
\begin{tabular}{|p{4.5cm}|p{2cm}|p{2cm}|p{2cm}|}
\hline
 parameter&mean for predictable days  &mean for not predictable days  &p-value  \\ \hline
Sample size (number of days) &14  &  66 &  \\ \hline
number of non-zero returns &2685 &  2454  &0.177  \\ \hline
fraction of 0-returns &0.100  &  0.094  &0.143  \\ \hline
$k$ &6 &  5.758  &7.594 $\times10^{-5}$** \\ \hline
$(\hat{p}(0\dots0)+\hat{p}(1\dots1))2^k$ &3.033 &2.462  &0.013*  \\ \hline
$|\hat{p}(1)-\hat{p}(0)|$ &0.068  &0.039 & 0.003**  \\ \hline
magnitude of daily log-price increment  &0.015  & 0.014 &0.952 \\ \hline
mean price returns  &-3.124$\times10^{-7}$ &-6.043$\times10^{-7}$ &0.874  \\ \hline
magnitude of autocorrelation of non-zero returns & 0.0299 & 0.0297 &0.966  \\ \hline
magnitude of autocorrelation of absolute values& 0.079  & 0.093 &0.208 \\ \hline
$\nu$ of t-distribution &2.656  & 3.426 &0.435\\ \hline
scale of t-distribution &1.843$\times10^{-4}$  & 2.156$\times10^{-4}$  &0.04* \\ \hline
magnitude of shift of t-distribution &1.609$\times10^{-5}$  & 1.204$\times10^{-5}$  &0.11  \\ \hline
daily volume  &12,788,633  & 12,048,587&0.438  \\ \hline
fraction of jumps  &2.093$\times10^{-4}$ & 1.685$\times10^{-4}$ &0.682  \\ \hline
\end{tabular}
\caption*{\footnotesize In the last column, * is rejection of equal means with 0.05 significance and ** stands for 0.01 significance.}
\label{Table: parameters AAPL with B}
\end{table}

\begin{table}[htb]
\centering
\caption{\footnotesize Statistics for predictable and not predictable days of CCL with $a=1$.}
\footnotesize
\begin{tabular}{|p{4.5cm}|p{2cm}|p{2cm}|p{2cm}|}
\hline
 parameter&mean for predictable days  &mean for not predictable days  &p-value  \\ \hline
Sample size (number of days) &38  &  42 &  \\ \hline
number of non-zero returns &2626 &  2195  &0.018*  \\ \hline
fraction of 0-returns &0.695 &  0.660 &0.0007**  \\ \hline
$k$ &5.816 &    5.595  &0.04*\\ \hline
$(\hat{p}(0\dots0)+\hat{p}(1\dots1))2^k$ &2.5 &2.25  &0.118  \\ \hline
$|\hat{p}(1)-\hat{p}(0)|$ &0.0205  & 0.0203  &0.956 \\ \hline
magnitude of daily log-price increment  &0.037 &  0.027 &0.1\\ \hline
mean price returns  &1.735$\times10^{-6}$ &  -1.491$\times10^{-6}$ & 0.016*  \\ \hline
magnitude of autocorrelation of non-zero returns &0.132&0.081&0.001** \\ \hline
magnitude of autocorrelation of absolute values& 0.286  &   0.234 &0.001**\\ \hline
$\nu$ of t-distribution &1.816  & 1.356 &0.002**\\ \hline
scale of t-distribution &1.281$\times10^{-5}$  &  2.868$\times10^{-5}$  &0.226\\ \hline
magnitude of shift of t-distribution &1.672$\times10^{-7}$  &  4.625$\times10^{-7}$  &0.194  \\ \hline
daily volume  &7,087,714 & 4,776,277&0.0001**  \\ \hline
fraction of jumps  &0.018& 0.016 &0.049* \\ \hline
\end{tabular}
\caption*{\footnotesize In the last column, * is rejection of equal means with 0.05 significance and ** stands for 0.01 significance.}
\label{Table: parameters CCL}
\end{table}

\begin{table}[htb]
\centering
\caption{\footnotesize Statistics for predictable and not predictable days of SPY with $a=5$.}
\footnotesize
\begin{tabular}{|p{4.5cm}|p{2cm}|p{2cm}|p{2cm}|}
\hline
 parameter&mean for predictable days  &mean for not predictable days  &p-value  \\ \hline
Sample size (number of days) &29  &  51 &  \\ \hline
number of non-zero returns &12764 &  10778  &0.04*  \\ \hline
fraction of 0-returns &0.170 &  0.168 &0.669  \\ \hline
$k$ &6.862 &    6.745  &0.217\\ \hline
$(\hat{p}(0\dots0)+\hat{p}(1\dots1))2^k$ &2.881 &2.539  &8.075$\times10^{-5}$** \\ \hline
$|\hat{p}(1)-\hat{p}(0)|$ &0.015  & 0.013  &0.417 \\ \hline
magnitude of daily log-price increment  &0.011 &  0.008&0.214\\ \hline
mean price returns  &-1.464$\times10^{-7}$ &  1.080$\times10^{-7}$ & 0.196 \\ \hline
magnitude of autocorrelation of non-zero returns &0.026& 0.018&0.009** \\ \hline
magnitude of autocorrelation of absolute values& 0.086  &   0.089 &0.76\\ \hline
$\nu$ of t-distribution &1.988  & 2.245 &0.031*\\ \hline
scale of t-distribution &5.531$\times10^{-5}$  &  5.467$\times10^{-5}$  &0.792\\ \hline
magnitude of shift of t-distribution &9.444$\times10^{-7}$  &  8.551$\times10^{-7}$  &0.544  \\ \hline
daily volume  &10,155,452&  8,554,678&0.037*  \\ \hline
fraction of jumps  &4.187$\times10^{-4}$& 5.052$\times10^{-4}$ &0.147 \\ \hline
\end{tabular}
\caption*{\footnotesize In the last column, * is rejection of equal means with 0.05 significance and ** stands for 0.01 significance.}
\label{Table: parameters SPY}
\end{table}

\begin{table}[htb]
\centering
\caption{\footnotesize Partition of predictable days}
\footnotesize
\begin{tabular}{|c|c|c|c|c|}
\hline
stock & day   &$S_{max}$& $S$  & N. $\in [1, S]$ \\ \hline
\multirow{2}{*}{{AAPL($a=30$)}} & 02.08, 05.10  &2& 1  & 1  \\ \cline{2-5} 
&20.09   &3& 1  & 1  \\ \hline
\multirow{4}{*}{{MSFT($a=15$)}}
 &03.08  &4& 4  & 4  \\ \cline{2-5} 
 &05.08  &3& 2  & 1  \\ \cline{2-5} 
 &26.10  &8& 2  & 2  \\ \cline{2-5} 
 &28.10  &4& 1  & 1  \\ \hline
\multirow{2}{*}{{TSLA($a=65$)}}
&05.08  &2& 2  & \textbf{1,2}  \\ \cline{2-5} 
&14.09,15.09, 08.11   &1& 1  & 1  \\ \hline
\multirow{21}{*}{{SNAP($a=1$)}}
&23.09,26.09,27.09,03.11  &2& 1  & 1  \\ \cline{2-5}
&15.09,08.11&2& 2  & 1  \\\cline{2-5}
&03.08,05.08,08.08,12.08,22.08, 14.09,03.10,12.10,28.10,14.11  &2& 2  & \textbf{1,2}  \\\cline{2-5}
&01.08,04.08,09.08,10.08,19.08, 12.09,28.09,17.10,31.10  &2& 2  & 2  \\\cline{2-5}
&09.09,15.11  &3& 1  & 1  \\ \cline{2-5}
&21.09  &3& 3  & 1  \\ \cline{2-5}
&06.09  &3& 3  & \textbf{1,2}  \\\cline{2-5}
&16.08,10.11,11.11  &3& 3  & 1,3  \\ \cline{2-5}
&18.08  &3& 3  & \textbf{2,3}  \\ \cline{2-5}
&02.08,27.10  &3& 3  & \textbf{1,2,3}  \\ \cline{2-5}
&15.08,06.10  &3& 3  & 3  \\\cline{2-5}
&11.08  &4& 2  & 1  \\\cline{2-5}
&17.08  &4& 3  & 2  \\ \cline{2-5}
&01.11  &4& 4  & \textbf{3,4}  \\ \cline{2-5}
&07.09,26.10  &4& 4  & 4  \\ \cline{2-5}
&08.09&5& 5  & 2  \\ \cline{2-5}
&20.10 &5& 5  & \textbf{2,3,4,5}  \\ \cline{2-5}
&25.10 &5& 5  & 3,5  \\ \cline{2-5}
&01.09 &5& 5  & \textbf{4,5}  \\ \cline{2-5}
&21.10,24.10 &7& 7  & \textbf{1,2,3,4,5,6,7} \\ \cline{2-5}
&31.08 &10& 10  & 1,\textbf{4,5},10 \\ \hline
\multirow{7}{*}{{SPY($a=10$)}}
&17.08  &5& 5  & 3  \\ \cline{2-5}
&30.08  &6& 1  & 1  \\ \cline{2-5}
&26.08,18.10  &7& 1  & 1  \\ \cline{2-5}
&01.09,02.09  &7& 2  & 2  \\\cline{2-5}
&29.09  &8& 1  & 1  \\ \cline{2-5}
&10.11 &9& 1  & 1  \\ \cline{2-5}
&13.10  &14& 10  & 10  \\ \hline
\end{tabular}
\caption*{\footnotesize Consecutive intervals for one day are in bold. For the stock SNAP, the results where days with $S_{max}=1$ are omitted.}
\label{optimal partition all}
\end{table}

\clearpage
\bibliographystyle{unsrt}
\bibliography{main}

\end{document}